\documentclass[]{aa}  

\usepackage{graphicx}
\begin{document}
\title{The contribution of Oxygen--Neon white dwarfs to the MACHO content of
       the Galactic Halo}

\titlerunning{ONe white dwarfs and the Galactic halo} 
\authorrunning{J. Camacho et al. } 

\author{Judit Camacho\inst{1},
        Santiago Torres\inst{1,2},
        Jordi Isern\inst{3,2},
        Leandro G. Althaus\inst{4,5}, 
        \and 
        Enrique  Garc\'{\i}a--Berro\inst{1,2}}

\institute{Departament de F\'\i sica Aplicada,   
          Escola Polit\'ecnica Superior de Castelldefels, 
          Universitat Polit\`ecnica de Catalunya,  
          Avda. del Canal Ol\'\i mpic s/n,  
          08860 Castelldefels, Spain\
          \and 
          Institute for Space Studies of Catalonia,
          c/Gran  Capit\`a 2--4, Edif. Nexus 104, 
          08034  Barcelona, Spain\
          \and
          Institut de Ci\`encies de l'Espai, CSIC,  
          Campus UAB, Facultat de Ci\`encies, Torre C-5, 
          08193 Bellaterra, Spain
          \and 
          Facultad de Ciencias Astron\'omicas y Geof\'{\i}sicas, 
          Universidad Nacional de La Plata, 
          Paseo del Bosque s/n, 1900 La Plata, Argentina\
          \and
          Instituto de Astrof\'{\i}sica La Plata, IALP, CONICET
          }
        
\offprints{E. Garc\'\i a--Berro}
\date{\today}

\abstract{The interpretation of microlensing results towards the Large
Magellanic Cloud (LMC) still remains controversial.  White dwarfs have
been  proposed to  explain  these results  and,  hence, to  contribute
significantly  to the  mass  budget of  our  Galaxy. However,  several
constraints on the role  played by regular carbon--oxygen white dwarfs
exist.}  {Massive white dwarfs are thought  to be made of a mixture of
oxygen and  neon. Correspondingly, their  cooling rate is  larger than
those  of  typical  carbon--oxygen  white  dwarfs  and  they  fade  to
invisibility  in short  timescales.  Consequently,  they  constitute a
good candidate  for explaining  the microlensing results.}   {Here, we
examine  in  detail this  hypothesis  by  using  the most  recent  and
up--to--date cooling tracks for massive white dwarfs and a Monte Carlo
simulator which takes into account the most relevant Galactic inputs.}
{We  find  that  oxygen--neon   white  dwarfs  cannot  account  for  a
substantial  fraction  of  the  microlensing depth  towards  the  LMC,
independently  of the  adopted  initial mass  function, although  some
microlensing events  could be  due to oxygen--neon  white dwarfs.}{The
white dwarf  population contributes at most  a 5\% to the  mass of the
Galactic halo.}

\keywords{stars:  white dwarfs  --- stars:  luminosity  function, mass
          function --- Galaxy: stellar content --- Galaxy: dark matter
          --- Galaxy: structure --- Galaxy: halo} 

\maketitle


\section{Introduction}

Several  cosmological  observations   show  compelling  evidence  that
baryons represent  only a  small fraction of  the total matter  in our
Universe and  that non--baryonic  dark matter dominates  over baryons.
To    be    specific,    in    the   standard    cosmological    model
$\Omega_{\Lambda}\simeq 0.72$ and $\Omega_{\rm M}\simeq 0.27$, whereas
$\Omega_{\rm  B}\simeq  0.044$.  Moreover,  most  of  the baryons  are
non--luminous, since $\Omega_\star\simeq 0.005$.   For the case of our
own Galaxy it  has been found that  the virial mass out to  100 kpc is
$M\approx 10^{12} \, M_{\sun}$ while  the baryonic mass in the form of
stars is $M_{\star} \approx  7 \times 10^{10}\, M_{\sun}$, which means
that for the Milky Way, the  baryon fraction is at most 8\% (Klypin et
al.  2007).   This problem is known  as the missing  bayon problem ---
see the  excellent review of  Silk (2007) for a  complete, interesting
and recent  discussion of  this issue  --- and it  is critical  in our
understanding of how the Galaxy  (an by extension other galaxies) were
formed and  will ultimately evolve.   In order to solve  this problem,
three alternatives can  be envisaged: either these baryons  are in the
outer regions of  our Galaxy, or, perhaps, they  never were present in
the protogalaxy or, finally, they may have been ejected from the Milky
Way. The most  promising explanation and the currently  favored one is
the first of these options. 

The most  likely candidates for  building up the baryonic  dark matter
density are  massive baryonic  halo objects, or  MACHOs.  It  has been
suggested that  MACHOs could be planets  ($M\sim 10^{-7}\, M_{\sun}$),
brown  dwarfs (with  masses ranging  from $\sim  0.01$ to  $\sim 0.1\,
M_{\sun}$),  primordial  black holes  ($M  \ga 10^{-16}\,  M_{\sun}$),
molecular clumps  ($M\sim 1\, M_{\sun}$) and old  white dwarfs ($M\sim
0.6 \, M_{\sun}$).  White  dwarfs are specially interesting candidates
not  only  because their  intrinsic  faintness,  but  also because  in
addition to the mass of  the white dwarf itself their progenitors have
to return to interstellar medium a sizeable fraction of their original
mass ($\sim 2\, M_{\sun}$ on  average) once the white dwarf is formed.
Additionally, much expectation has been generated since the pioneering
proposal  of Paczy\'nski  (1986) that  MACHOs could  be  found through
gravitational microlensing.   Since then,  several groups such  as the
MACHO  (Alcock et  al.   1997,  2000), EROS  (Lasserre  et al.   2001,
Goldman et  al.  2002, Tisserand et  al.  2006), OGLE  (Udalski et al.
1994), MOA (Muraki et al.   1999) and SuperMACHO (Becker et al.  2005)
teams have  monitored millions of  stars during several years  in both
the Large Magellanic Cloud (LMC)  and the Small Magellanic Cloud (SMC)
to search for microlensing events.   Among these searches, it is worth
mentioning  that the  MACHO collaboration  has succeeded  in revealing
13--17  microlensing  events during  their  5.7  yr  analysis of  11.9
million stars  in the  LMC (Alcock et  al.  2000).  In  their analysis
they    derived    an   optical    depth    towards    the   LMC    of
$\tau=1.2^{+0.4}_{-0.3}\times  10^{-7}$ for  events with  durations in
the  range $2<\hat{t}<400\ {\rm  days}$.  This  value is  smaller than
that expected  for a full  MACHO halo.  In  fact, it corresponds  to a
halo fraction $0.08<f<0.50$ at the  95\% confidence level with a MACHO
mass  in  the  range  $0.15\,  M_{\sun}\leq  M  \leq  0.50\,M_{\sun}$,
depending on the halo model.  Despite the fact that only a fraction of
the dark matter could be in the form of MACHOs, there is still a large
controversy about  the nature of the reported  microlensing events and
to which extent they contribute to the mass budget of the dark halo of
the Galaxy.   In fact, a  large variety of possible  explanations have
been  proposed to  explain these  microlensing events.   For instance,
white  dwarfs,  brown  dwarfs   and  black  holes  appear  as  natural
candidates,  whereas self--lensing  by stars  of the  LMC  (Sahu 1994,
Gyuk, Dalal  \& Griest 2000) has  been proposed as  well.  Also, other
explanations ---  like tidal debris or  a dwarf galaxy  toward the LMC
(Zhao  1998), a galactic  extended shroud  population of  white dwarfs
(Gates  \& Gyuk  2001), blending  effects (Belokurov,  Evans \&  Le Du
2003,  2004),  non--conventional  initial  mass  functions  (Adams  \&
Laughlin 1996; Chabrier et al. 1996), spatially varying mass functions
(Kerins  \&   Evans  1998,   Rahvar  2005),  and   other  explanations
(Holopainen  et al.   2006) ---  have been  also  thoroughly discussed
during  the last  years.  However,  all of  these proposals  have been
received with some  criticism because none of them  fully explains the
observed microlensing results.

There  are as  well other  observations that  are important  pieces of
evidence in this puzzle, such as the results of the EROS collaboration
or the  search for  very faint  objects in the  Hubble Deep  Field. We
briefly  summarize  them. The  EROS  team  has  recently presented  an
analysis of a  subsample of bright stars from  the LMC, minimizing the
source confunsion and blending  effects (Tisserand et al. 2006). Their
results   imply  that   the   optical  depth   towards   the  LMC   is
$\tau<0.36\times 10^{-7}$ at  the 95\% confidence level, corresponding
to a fraction of halo mass of  less than $7\%$. This result is 4 times
smaller than that obtained by the MACHO team and, consequently, sets a
strong upper bound to the contribution of MACHOs to the mass budget of
the  Galactic  dark matter  halo.   Nevertheless,  the  nature of  the
observed microlensing events still  remains to be clarified. Also, the
Hubble Deep Field--South has  provided another opportunity to test the
contribution  of  white  dwarfs  to  the  Galactic  dark  matter.   In
particular, Kilic et al. (2005)  have recently found three white dwarf
candidates among several faint  blue objects which exhibit significant
proper motion and,  thus, are assumed to belong  to the thick--disk or
the  halo  populations.   If  in   the  end  these  white  dwarfs  are
spectroscopically  confirmed  it would  imply  that  white dwarfs  can
account for about $\la 10\%$  of the Galactic dark matter, which would
be consistent  with the  results of the  EROS team, and  with previous
estimates (Chabrier 2004).

In a  previous paper (Garc\'{\i}a--Berro  et al. 2004)  we extensively
analyzed  the  role played  by  the  carbon--oxygen  (CO) white  dwarf
population  in several  different observational  results,  namely, the
reported  microlensing  events  towards  the  Large  Magellanic  Cloud
(Alcock et al.  2000), the results  of the Hubble Deep Field (Ibata et
al.   1999)  and  the  results  of the  EROS  experiment  (Goldman  et
al. 2002). We performed a thorough  study for a wide range of Galactic
inputs, including different initial  mass functions and halo ages, and
several density profiles corresponding  to different halo models.  Our
main  result was  that a  sizeable fraction  of the  halo  dark matter
cannot be locked  in the form of old  hydrogen--rich white dwarfs with
CO cores. Specifically,  we found that this fraction  should be of the
order of $4\%$, in agreement  with the standard models of the Galactic
halo.   However in  our analysis  we disregarded  the  contribution of
massive white  dwarfs, that  is, stars more  massive than  $\sim 1.1\,
M_{\sun}$. The  core of  these white dwarfs  consists of a  mixture of
oxygen  and   neon.   Since  oxygen--neon  (ONe)   white  dwarfs  cool
considerably faster than  the bulk of CO white  dwarfs (Althaus et al.
2007) it is reasonable to expect that perhaps some of the microlensing
events could be due to these elusive massive white dwarfs.  It is also
worth mentioning at  this point that the MACHO  collaboration in their
first season reported a microlensing event with a duration of 110 days
towards the galactic bulge (Alcock et al.  1995).  For this particular
event a parallax could be obtained  from the shape of the light curve,
from  which  a  mass  of  $1.3^{+1.3}_{-0.6}\,M_{\sun}$  was  derived,
indicating that the gravitational lens could possibly be a massive ONe
white  dwarf   or  a  neutron  star.   Moreover,   studies  about  the
distribution of  masses of the  white dwarf population (Finley  et al.
1997; Liebert et al.  2005) show  the existence of a narrow sharp peak
near  $0.6\,M_{\sun}$, with  a tail  extending towards  larger masses,
with  several white  dwarfs with  spectroscopically  determined masses
within the interval comprised between $1.0$ and $1.2\,M_{\sun}$.

In this paper we analyze if  ONe white dwarfs could be responsible for
a sizeable  fraction of the  reported microlensing events  towards the
LMC.   The paper  is organized  as follows.   In Sect.   2  we briefly
describe the main ingredients of  our Monte Carlo code and other basic
assumptions  and  procedures necessary  to  evaluate the  microlensing
optical depth  towards the LMC. Section  3 is devoted  to describe our
main results,  including the contribution  of ONe white dwarfs  to the
halo white  dwarf luminosity function and to  the microlensing optical
depth towards  the LMC,  and we  compare our results  to those  of the
MACHO  and EROS  teams. In  this section  we also  check if  ONe white
dwarfs could be detected in the Hubble Deep Field South and we discuss
the contribution  of ONe white dwarfs  to the baryonic  content of the
Galaxy. Finally, in  Sect. 4 our major findings  are summarized and we
draw our conclusions.

\section{The model}

\subsection{The Monte Carlo simulation}

An extensive description of our Monte Carlo simulator has been already
presented in Garc\'{\i}a--Berro et  al.  (2004). Consequently, here we
will only  briefly summarize  the main ingredients  of our  model.  We
have  used a  random  number generator  algorithm  (James 1990)  which
provides a uniform probability density within the interval $(0,1)$ and
ensures a repetition period of  $\ga 10^{18}$, which is enough for our
purposes.   Each  one of  the  Monte  Carlo  simulations discussed  in
Sect. 3 below  consists of an ensemble of  40 independent realizations
of the synthetic white dwarf  population, for which the average of any
observational quantity along with its corresponding standard deviation
were computed.  Here the standard deviation means the ensemble mean of
the sample dispersions for a typical sample.

We have  considered an  otherwise typical spherically  symmetric halo.
The density profile of this model is the isothermal sphere of radius 5
kpc, also called the ``S--model'',  which has been extensively used by
the MACHO  collaboration (Alcock et al.  2000;  Griest 1991).  Despite
the  existence of  other  density profiles,  such  as the  exponential
power--law model, the Navarro,  French \& White (1997) density profile
and others, in  our previous study (Garc\'\i a--Berro  et al. 2004) we
showed that the  differences between them are not  significant for the
case  under  study  and,   consequently,  we  adopt  the  most  simple
description. The  position of each  synthetic star is  randomly chosen
according to this density profile.

We have  considered two different  initial mass functions,  the rather
standard  initial  mass  function  of  Scalo  (1998)  and  the  biased
log--normal  initial  mass  function  proposed by  Adams  \&  Laughlin
(1996), which  is very similar  to the non--conventional  initial mass
function of Chabrier et al. (1996).  This biased initial mass function
has been included just for the sake of completeness, since it does not
seem to be  compatible with the observed properties  of the halo white
dwarf population (Isern et al.  1998; Garc\'\i a--Berro et al.  2004),
with the  contribution of thermonuclear supernovae  to the metallicity
of the Galactic halo (Canal et al. 1997), and with the observations of
galactic halos in deep galaxy surveys (Charlot \& Silk 1995). The main
sequence mass is obtained by drawing a pseudo--random number according
to the  adopted IMF.   Once the  mass of the  progenitor of  the white
dwarf is  known we  randomly choose  the time at  which each  star was
born.  We  assume that the  halo was formed  14 Gyr ago in  an intense
burst of  star formation of duration  $\sim 1$~Gyr.  Given  the age of
the halo,  the time at  which each main--sequence progenitor  was born
and the main  sequence lifetime as a function of the  mass in the main
sequence (Iben  \& Laughlin  1989) we know  which stars have  had time
enough to enter  in the white dwarf cooling track, and  given a set of
theoretical  cooling   sequences  and   the  initial  to   final  mass
relationship  (Iben \&  Laughlin  1989), we  know their  luminosities,
effective temperatures and colors.  The cooling sequences adopted here
depend  on the  mass of  the white  dwarf.  White  dwarfs  with masses
smaller than  $M_{\rm WD}=1.1\,  M_{\sun}$ are expected  to have  a CO
core  and, consequently,  for  them  we adopt  the  cooling tracks  of
Salaris et al.  (2000).  White  dwarfs with masses larger than $M_{\rm
WD}=1.1\, M_{\sun}$ most  probably have ONe cores and  for these white
dwarfs we  adopt the most recent  cooling sequences of  Althaus et al.
(2007).  Both sets of  cooling sequences incorporate the most accurate
physical  inputs  for   the  stellar  interior  (including  neutrinos,
crystallization, phase separation and Debye cooling) and reproduce the
blue turn  at low luminosities  (Hansen 1998).  Also, the  ensemble of
cooling sequences  used here encompass  the full range of  interest of
white dwarf masses, so a complete  coverage of the effects of the mass
spectrum of the white dwarf population was taken into account.

The  kinematical properties of  the halo  white dwarf  population have
been modeled according to a gaussian law (Binney \& Tremaine 1987):

\begin{equation}
f(v_r,v_t)=\frac{1}{(2\pi)^{3/2}}\frac{1}{\sigma_r\sigma_t^2}
\exp\left[-\frac{1}{2}\left(\frac{v_r^2}{\sigma_t^2}+\frac{v_{t}^2}
{\sigma_t^2}\right)\right]
\end{equation}

\noindent  where $\sigma_r$  and  $\sigma_t$ ---  the  radial and  the
tangential velocity  dispersion, respectively  --- are related  by the
following expression:

\begin{equation} 
\sigma_t^2=\frac{V_{\rm   c}^2}{2}+\left[1-\frac{r^2}{a^2+r^2}\right]
\sigma_r^2+\frac{r}{2}\frac{{\rm d}(\sigma_r^2)}{{\rm d}r}
\end{equation}

\noindent which  reproduces the flat  rotation curve of our  Galaxy at
large  distances. We  have  adopted a  circular  velocity $V_{\rm  c}=
220$~km/s. Finally, and in order to obtain the heliocentric velocities
we have take into account the peculiar velocity of the sun $(U_{\sun},
V_{\sun},W_{\sun})=(10.0,   15.0,   8.0)$~km/s   (Dehnen   \&   Binney
1998). Since  white dwarfs usually  do not have determinations  of the
radial component  of the velocity,  the radial velocity  is eliminated
when a comparison with the observational data is needed.  Moreover, we
only consider stars with velocities larger than 250~km/s because white
dwarfs with  velocities smaller than  this would not be  considered as
halo  members. Additionally,  we  also discard  stars with  velocities
larger than 750~km/s, because they would have velocities exceeding 1.5
times the escape velocity.

\subsection{Modeling the microlensing events towards the LMC}

In order  to ascertain  the contribution of  halo white dwarfs  to the
microlensing events towards the LMC  we have proceeded in three steps.
First of  all we have built a  model of the LMC  following closely the
procedures  detailed  in  Gyuk  et  al.  (2000)  and  Kallivayalil  et
al. (2006). This model takes into account, among other parameters, the
scale  length and scale  height of  the LMC,  its inclination  and its
kinematical  properties.   This model  provides  us  with a  synthetic
population of stars representative of the monitored point sources.  In
a second  step we  search for  those halo white  dwarfs that  could be
responsible  of a  microlensing event.   This implies  that  the white
dwarf should be fainter than a magnitude limit, otherwise it would not
be considered as a genuine microlensing event. Typically we have taken
$m_{\rm V}^{\rm cut}=  17.5^{\rm mag}$, which is the  value adopted by
Alcock  et  al.   (2000).  This  value  has  been  confirmed to  be  a
reasonable  estimate  by   the  detailed  theoretical  simulations  of
Garc\'\i a--Berro  et al.  (2004).   Finally, we check if  the angular
distance between  the white  dwarf and the  monitored star  is smaller
than the Einstein radius  $\theta_{\rm E}=R_{\rm E}/D_{\rm OL}$, where
$D_{\rm OL}$  is the  distance between the  observer and the  lens and
$R_{\rm E}$ is the Einstein radius which is given by the expression

\begin{equation}
R_{\rm E}=2\sqrt{\frac{GMD_{\rm OS}}{c^2}x(1-x)}
\end{equation}

\noindent  where $D_{\rm  OS}$  is the  observer--source distance  and
$x\equiv D_{\rm  OL}/D_{\rm OS}$. If this  is the case then  we have a
microlensing event and we  compute the corresponding probability. This
probability  is  integrated  over   the  total  monitoring  period  of
observation and  filtered by the detection  efficiency function, which
allows us ot obtain the optical depth (Alcock et al. 2000):

\begin{equation}
\tau=\frac{1}{E}\frac{\pi}{4}\sum_i  
\frac{\hat  t_i}{\varepsilon(\hat t_i)}
\end{equation}

\noindent where $E$ is the total exposure (in star--years), $\hat t_i$
is  the Einstein  ring diameter  crossing time,  and $\varepsilon(\hat
t_i)$ is  the detection efficiency.  The detection  efficiency and $E$
depend on the particular characteristics of the experiment and, hence,
we  consider  different  detection  efficiencies and  different  total
exposures for  the MACHO and EROS experiments.   Specifically, for the
case in  which we  analyze the results  of the MACHO  collaboration we
have taken $1.1\times  10^7$ stars during 5.7 yr  and over $13.4\ {\rm
deg^2}$, whereas the detection efficiency has been modeled as:

\begin{equation}
\varepsilon(\hat{t})=
\left\{
\begin{array}{cc}
0.43\,{\rm e}^{-(\ln(\hat{t}/T_{\rm m}))^{3.58}/ 0.87}, & \hat{t}>T_{\rm m} \\
0.43\,{\rm e}^{-|\ln(\hat{t}/T_{\rm m})|^{2.34}/11.16}, & \hat{t}<T_{\rm m} 
\end{array}
\right.
\end{equation}

\noindent where $T_{\rm m}=250$ days.  This expression provides a good
fit to the results of Alcock  et al.  (2000).  For the EROS experiment
we  have  used $0.7\times  10^7$  stars over  a  wider  field of  $84\
{\deg^2}$  and over  a period  of  $6.7$ yr.  Regarding the  detection
efficiency we have adopted a numerical fit to the results presented in
Tisserand et al. (2006).

\begin{figure}
\vspace{13cm}
\includegraphics{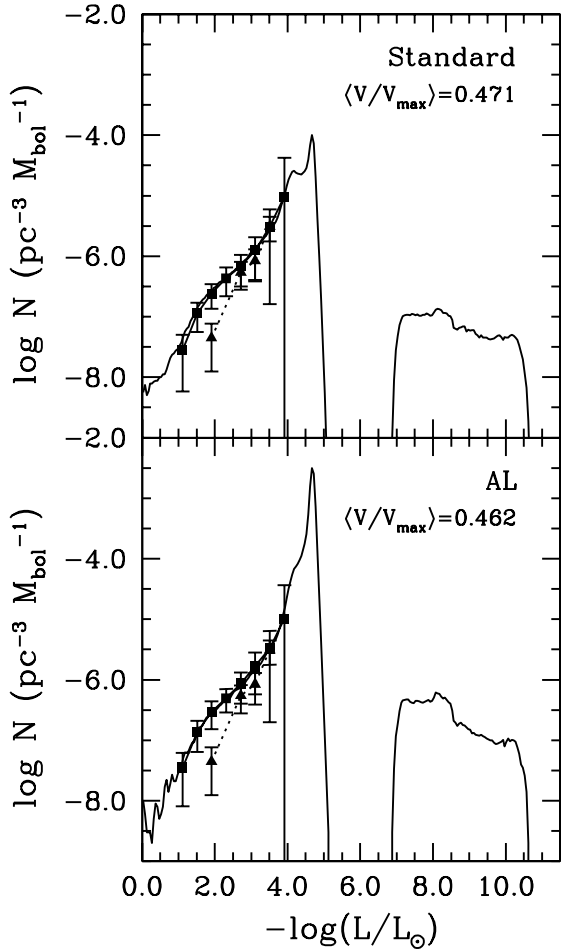}
\caption{Luminosity  function  of halo  white  dwarfs  for a  standard
         initial mass  function (top panel) and a  biased initial mass
         function   (bottom  panel).   The   observational  luminosity
         function of  halo white dwarfs is represented  using a dotted
         line  (Torres et  al. 1998)  and solid  triangles,  while the
         theoretical luminosity  function is shown using  a solid line
         and squares. See text for details.}
\end{figure}

\begin{table*}
\centering
\begin{tabular}{lrrrrrrrr}
\hline      
\hline      
\multicolumn{1}{c}{\ }           &      
\multicolumn{4}{c}{Standard}     &   
\multicolumn{4}{c}{AL}           \\
\hline 
Magnitude  &  17.5 & 22.5 & 27.5 & 32.5   
           &  17.5 & 22.5 & 27.5 & 32.5 \\  
\cline{2-5}  
\cline{6-9}    
$\langle N_{\rm  WD}\rangle$ &  
$0\pm 1$ &  $0\pm 1$ &  $0\pm 1$ & $0\pm  1$ & $3\pm  3$ & $2\pm  2$  &   $1\pm  1$  &  $0\pm  2$  \\   
$\langle m \rangle$ $(M/M_{\sun})$ &  
0.593 &  0.599 & 0.619 & 0.888 & 0.636 & 0.638 & 0.651 & 0.684 \\  
$\langle \mu\rangle$  $(''\,{\rm yr}^{-1})$  &
0.018 &  0.015 & 0.009 & 0.004 & 0.038 & 0.025 & 0.010 & 0.003 \\
$\langle d\rangle$ (kpc) & 
2.85  &  3.52  & 6.27  & 14.65 & 1.31  & 2.22  & 5.45  & 18.73 \\  
$\langle V_{\rm tan} \rangle$ $({\rm km\,s}^{-1})$ & 
238   &  243   & 262   & 268   &  240  & 260   & 257   & 279   \\ 
$\langle\hat{t}_{\rm E}\rangle $ (d) &  
56.6  &  59.8  & 82.4  & 121.2 &  34.9 & 48.0  & 76.6  & 129.7 \\ 
$\langle  \tau/\tau_0 \rangle$ &  
0.139 & 0.134  & 0.187 & 0.131 & 0.180 & 0.162 & 0.167 & 0.192 \\ 
\hline
\hline
\end{tabular}
\caption{Summary  of  the  results  obtained  for  the  simulation  of
	microlenses towards the LMC for  the MACHO model for an age of
	the  halo  of  14~Gyr,   different  model  IMFs,  and  several
	magnitude cuts.}
\end{table*}

\section{Results}

\subsection{The halo white dwarf luminosity function}

\begin{figure}
\vspace{13cm}
\includegraphics{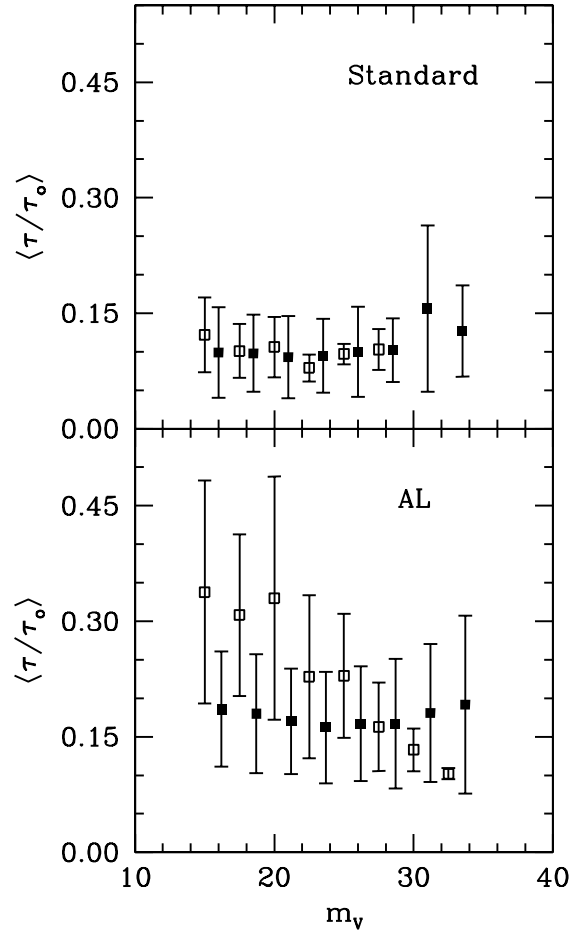}   
\caption{Microlensing optical  depth towards the LMC as  a function of
         the limiting magnitude.  Open and solid simbols represent the
         population of white dwarfs  without and with the contribution
         of  the ONe  white dwarfs,  respectively.  The  solid symbols
         have been shifted for the sake of clarity.}
\end{figure}

Despite the  increasing number of  surveys searching for  white dwarfs
--- like the Sloan Digital Sky Survey (Eisenstein et al.  2006), the 2
Micron All Sky Survey (Cutri  et al. 2003), the SuperCosmos Sky Survey
(Hambly,  Irwin \&  MacGillivray 2001),  the 2dF  QSO  Redshift Survey
(Vennes et  al.  2002), and others  --- their success  in finding halo
white dwarfs has been  limited.  Thus, the observational determination
of the  halo white dwarf  luminosity is still today  rather uncertain.
In fact, the two attempts to build such a luminosity function (Liebert
et al. 1989; Torres et al. 1998) have provided us only with the bright
branch  of the  halo white  dwarf luminosity  function.  Nevertheless,
this is  enough for our purposes,  since we only  need a normalization
criterion  and, hence, only  an upper  limit to  the local  density of
moderately  bright dwarfs is  needed. Consequently,  we have  used the
luminosity function  of Torres et  al.  (1998) and we  have normalized
the  local density  of  white  dwarfs obtained  from  our Monte  Carlo
simulations to its observed value, $n\sim 9.0\times 10^{-6}$~pc$^{-3}$
for $\log(L/ L_{\sun})\ga -3.5$ (Torres et al.  1998).

From the distribution  of white dwarfs obtained using  our Monte Carlo
simulations we compute the  white dwarf luminosity using the $1/V_{\rm
max}$ method  (Schmidt 1968).   It is important  to mention  that when
deriving  a luminosity  function using  the $1/V_{\rm  max}$  method a
proper motion cut  and a limiting magnitude are  required.  The set of
selection  criteria adopted here  for computing  the halo  white dwarf
luminosity function  is the same as  used in Garc\'\i  a--Berro et al.
(2004).  Namely,  we have chosen a limiting  magnitude $m_{\rm V}^{\rm
lim}=17.5^{\rm   mag}$   and   a    proper   motion   cut   $\mu   \ge
0.16^{\prime\prime}$~yr$^{-1}$.  With all  these inputs the luminosity
functions in Fig.  1 are obtained.  The top panel shows the halo white
dwarf  luminosity  function obtained  using  a  standard initial  mass
function, whereas the bottom  panel shows the luminosity function when
the  biased initial  mass  function  of Adams  \&  Laughlin (1996)  is
adopted.   The  simulated  luminosity  functions  are  represented  as
squares  connected   with  solid  lines,   whereas  the  observational
luminosity function is represented  as triangles connected with dashed
lines. We also recall that, by construction, our samples are complete,
although  we only  select about  10 white  dwarfs using  the selection
criteria discussed  before.  However,  our simulations do  provide the
whole population of white dwarfs,  which is much larger. Hence, we can
obtain  the {\sl real}  luminosity function  by simply  counting white
dwarfs in the computational volume.  This is done for all realizations
and then we obtain the average. The result is depicted as a solid line
in  Fig.  1.   The  true luminosity  function  steadily increases  for
luminosities larger than $\log(L/L_{\sun})\simeq-5.0$ and then sharply
drops. This drop--off is given by  the paucity of CO white dwarfs with
appropriate  ages  (14  Gyr).  Note  however  that  the  bulk  of  the
population   of  ONe  white   dwarfs  is   located  at   much  smaller
luminosities, a consequence of  the much shorter cooling timescales of
these white  dwarfs.  In fact, for a  typical halo age of  14 Gyr, the
bulk of  the ONe white dwarf  population has already  entered the fast
Debye cooling  phase and, consequently,  would not be  detectable with
the current observational facilities.  In the next sections we explore
if  this   elusive  white  dwarfs  contribute   significantly  to  the
microlensing optical depth. It is also important to note that with the
adopted limiting  magnitude and proper motion cut  we obtain simulated
white dwarf luminosity functions which are totally compatible with the
observational one.   Hence, the inclusion of massive  ONe white dwarfs
does  not  appreciably change  the  resulting  white dwarf  luminosity
function, which is very similar  to that obtained in Garc\'\i a--Berro
et al. (2004).

\subsection{Microlensing towards the LMC}

Firt  of   all,  we   analyze  the  result   obtained  by   the  MACHO
collaboration.  In  Fig.  2  we show the  contribution to  the optical
depth towards the LMC due to  the white dwarf population as a function
of the adopted limiting magnitude. The results have been normalized to
the value derived by Alcock et al. (2000), $\tau_0=1.2\times 10^{-7}$.
The open  symbols represent the  contribution if only CO  white dwarfs
are taken into account, while  the solid symbols show the contribution
to the  microlensing optical depth when  both CO and  ONe white dwarfs
are correctly included in the  model white dwarf population. As can be
seen, for none of the  adopted initial mass functions the inclusion of
the  ONe white dwarf  population does  not significantly  increase the
contribution of white dwarfs to the microlensing optical depth towards
the LMC, despite the fact that  ONe white dwarfs are much fainter than
regular  CO  white  dwarfs  (see  also  Fig.   1).  Specifically,  the
contribution of  the white dwarf  population is, respectively,  of the
order of 10\%  for the case of the standard  initial mass function and
somewhat  larger  ($\sim  15\%$)  for  the  log--normal  initial  mass
function of Adams \& Laughlin (1996).  These figures are comparable to
those already  found in  Garc\'\i a--Berro et  al.  (2004).   The only
differences are  that in  the case of  the standard mass  function the
contribution of ONe white dwarfs  to the microlensing optical depth is
clearly dominant  only when the  adopted limiting magnitude is  of the
order of  30, which is a  totally unrealistic value.  For  the case of
the log--normal initial mass  function the results presented here show
that the contribution is nearly constant, independently of the adopted
limiting  magnitude, whereas when  only the  contribution of  CO white
dwarfs was  considered the  contribution to the  optical depth  of the
halo  white dwarf  population  was clearly  decreasing for  increasing
magnitude cuts.

A summary of  the results obtained with our  Monte Carlo simulator can
be found  in Table 1, where  we show for four  selected magnitude cuts
the  number   of  microlensing  events,   the  average  mass   of  the
microlenses,  their  average proper  motion,  distance and  tangential
velocity, the corresponding Einstein  crossing times and, finally, the
contribution to  the microlensing optical  depth.  It is  important to
discuss some of  the numerical values in Table 1.  For instance, it is
clear that the larger the  magnitude cut, the more massive the average
mass  of  the lenses,  as  it  should be  expected  from  Fig.  1.  In
particular, for the case in  which a standard initial mass function is
used we  obtain that  for the largest  limiting magnitude  the average
mass is $\sim 0.9\, M_{\sun}$, indicating that in sizeable fraction of
the Monte  Carlo realizations the lens  is an ONe  white dwarf.  Also,
the  log--normal  initial  mass  function produces  more  microlensing
events,  as one  should expect,  given that  this biased  initial mass
funtion was tailored to produce  more microlensing events. In fact for
this initial mass  function a maximum number of  6 microlensing events
should be  expected, while for  the standard initial mass  function we
should   expect  1   microlensing  event,   at  most.    However,  the
contribution to the microlensing optical depth is only slightly larger
for the  Adams \& Laughlin  (1996) initial mass function.   The reason
for this  is that the  microlensing events for this  distribution have
shorter Eintein crossing times, as seen in table~1.

\begin{table*}
\centering
\begin{tabular}{lrrrrrrrr}
\hline
\hline
\multicolumn{1}{c}{\ }         &
\multicolumn{4}{c}{Standard}   &
\multicolumn{4}{c}{AL}\\
\hline 
Magnitude &  17.5 & 22.5 & 27.5 & 32.5 
          &  17.5 & 22.5 & 27.5 & 32.5  \\ 
\cline{2-5}  
\cline{6-9}   
$\langle m\rangle$ $(M/M_{\sun})$ &  
1.118 &  1.106 & 1.244 & 1.130 & 1.092 & 1.082 & 1.083 & 1.101 \\
$\langle d\rangle$ (kpc) & 
4.95  &  3.98  & 2.83  & 3.80  & 2.31  & 2.02  & 2.84  & 5.75  \\ 
$\langle V_{\rm tan} \rangle$ $({\rm km\,s}^{-1})$ & 
253   &  257   & 250   & 250   & 266   & 255   & 250   & 269   \\ 
$\langle \hat{t}_{\rm E}\rangle $ (d)  &  
107.9 & 91.6   & 77.7  & 104.1 & 56.8  & 61.6  & 75.0  & 99.0  \\ 
\hline 
\hline
\end{tabular}
\caption{Average values for the ONe white dwarf population.}
\end{table*}

\begin{figure}
\vspace{6cm}
\includegraphics{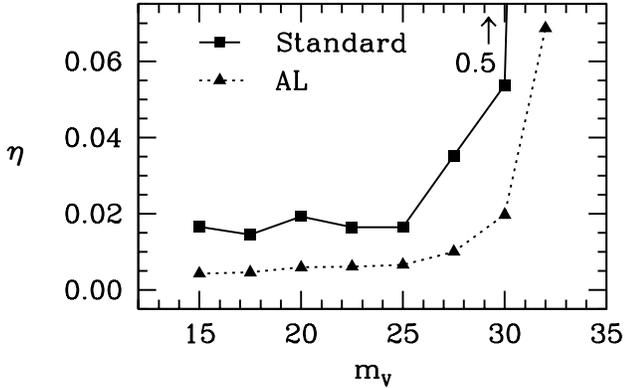}
\caption{Fraction of microlenses due  to ONe white dwarfs with respect
         to  the whole  population of  white dwarfs  for  the standard
         initial mass function --- squares --- and for the log--normal
         initial  mass  function  of  Adams  \&  Laughlin  (1996)  ---
         triangles.}
\end{figure}

The results obtained so far are not evident at first glance, since one
may  expect  that  ONe   white  dwarfs  should  be  good  microlensing
candidates. As  previously mentioned, ONe  white dwarfs have  a faster
cooling  rate than  that of  CO white  dwarfs and,  consequently, they
reach much  fainter magnitudes for  the same cooling age.   Hence, for
reasonable halo  ages one should  naively expect that  the probability
that a  ONe white  dwarf could produce  a microlensing event  would be
somewhat  larger  than  that of  a  CO  white  dwarf, given  that  for
reasonable halo ages practically  all ONe white dwarfs have magnitudes
larger than the magnitude cuts  adopted here. However, even if this is
indeed  the case, we  have shown  that the  total contribution  of ONe
white dwarfs  is almost  negligible.  To clarify  this result  we have
analyzed  the fraction  of microlenses  due to  ONe white  dwarfs with
respect to  that of  the total  population.  In Fig.   3 we  show this
fraction as a  function of the limiting magnitude  for the two initial
mass functions under  study.  As can be seen,  the contribution of ONe
white dwarfs  is small for  limiting magnitudes below  $25^{\rm mag}$.
Specifically, for the case of  the standard initial mass function they
only contribute  a modest $2\%$,  whereas for the  log--normal initial
mass function the contribution is halved. This situation only reverses
when magnitude cuts larger than $\sim 27^{\rm mag}$ are adopted.  This
result by  itself is  not explanatory of  why the contribution  of ONe
white dwarfs is not significant.  We recall here that the contribution
of an  object to the  total optical depth  is given by Eq.  (4), which
depends on the  Einstein crossing time which, in  turn, depends on the
Einstein radius  and on the  transverse velocity of the  lens, $t_{\rm
E}=r_{\rm E}/v_{\rm tan}$.  The Einstein  radius scales as the root of
the  mass  of the  object  and it  also  depends  on the  lens--object
distance ---  see Eq.  (3).  We note  that the average mass  of an ONe
white dwarf is larger than that  of a CO white dwarf.  Also, given the
intrinsic faintness of ONe white dwarfs, their spatial distribution in
the  computational  volume  is  different  because  we  are  selecting
microlensing candidates with magnitudes  fainter than a given limiting
magnitude.   Thus, it  can be  expected that  the contribution  to the
optical depth of an representative object of the these two populations
should be different as well.
 
In  Table 2  we show  the average  parameters of  the ONe  white dwarf
population susceptible  to produce a microlensing  event.  The average
mass of an ONe white dwarf  is $\simeq 1.1\, M_{\sun}$, while for a CO
white  dwarf it is  $\simeq 0.6\,  M_{\sun}$. On  the other  hand, the
average distance of  ONe white dwarfs is in the  range between about 2
and 4 kpc,  independently of the limiting magnitude,  while for the CO
white dwarf  population the average distance  increases for increasing
magnitude cuts.   Finally, the average tangential  velocities are very
similar for all the magnitude  cuts, given that the selection criteria
are  independent of  the kinematical  properties of  the  sample. With
these data and using Eq. (3)  and (4) the ratio of the contribution to
the optical  depth of a  typical ONe white  dwarf with respect  to the
contribution of a typical CO white dwarf is

\begin{equation}
\frac{\tau_{\rm ONe}}{\tau_{\rm CO}}=
\frac{\hat t_{\rm ONe}}{\hat t_{\rm CO}}
\frac{\varepsilon(\hat t_{\rm CO})}{\varepsilon(\hat t_{\rm ONe})}
\approx\sqrt{\frac{M_{\rm ONe}D_{\rm OL}^{\rm ONe}}
{M_{\rm CO}{D_{\rm OL}^{\rm CO}}}}
\frac{\varepsilon(\hat t_{\rm CO})}{\varepsilon(\hat t_{\rm ONe})}
\end{equation}

This  ratio  turns  out  to be  $\tau_{\rm  ONe}/\tau_{\rm  CO}\approx
1.5$.  Recalling that  the fraction  $\eta$  of ONe  white dwarfs  for
limiting magnitudes  fainter than  $25^{\rm mag}$ is  typically $0.02$
for  the standard  initial mass  function  and $0.01$  for the  biased
initial mass function, the increment in the total optical depth due to
ONe white dwarfs can be estimated to be

\begin{equation}
\frac{\Delta\tau}{\tau_0}\approx\eta\frac{\tau_{\rm ONe}}{\tau_{\rm CO}},
\end{equation}

\noindent which represents an increment  of roughly $3\%$ for the case
in which  a standard initial mass  function is considered  and a $2\%$
increment for the case of the log--normal initial mass function. These
results are in nice agreement  with those previously presented in Fig.
2.  On the  other hand, when the  magnitude cut is  $30^{\rm mag}$ the
fraction of ONe microlenses  $\eta$ increases significantly and, thus,
the fractional increase  of the optical depth due  to ONe white dwarfs
consequently increases, reaching values  as high as $100\%$. This fact
is  responsible for the different  behaviour of the  deepest magnitude
bins of  the left panel of Fig.   2, which show the  situation for the
standard  initial  mass function.  The  biased  initial mass  function
suppresses the  formation of moderately massive ONe  white dwarfs, and
this  is the  reason why  these faintest  luminosity bins  are  not as
populated as the equivalent bins for the case in which a standard mass
function is considered.

\begin{table*}
\centering
\begin{tabular}{lrrrrrr}
\hline
\hline
\multicolumn{1}{c}{\ }       &
\multicolumn{3}{c}{Standard} &   
\multicolumn{3}{c}{AL}       \\
\hline 
Magnitude  & 17.5 &  22.5 & 27.5  
           & 17.5 &  22.5 & 27.5 \\ 
\cline{2-4}  
\cline{5-7}  
$\langle N_{\rm WD}\rangle$ & 
$0\pm 1$ & $0\pm 1$ & $0\pm 1$ & $1\pm 2$ & $1\pm 2$ & $0\pm 2$ \\
$\langle m \rangle$  $(M/M_{\sun})$ & 
0.607    & 0.595    & 0.622    & 0.631    & 0.634    & 0.642    \\ 
 $\langle \mu\rangle$  $(''\,{\rm yr}^{-1})$  &
0.013    & 0.011    & 0.008    & 0.034    & 0.025    & 0.010    \\ 
$\langle d\rangle$ (kpc) & 
4.29     & 4.52     & 6.71     & 1.50     & 2.03     & 5.39     \\
$\langle V_{\rm tan} \rangle$ $({\rm km\,s}^{-1})$ & 
256      & 239      & 246      &  240     & 244      & 258      \\ 
$\langle \hat{t}_{\rm E}\rangle $ (d) &  
64.9     & 77.0     & 89.7     & 37.9     &  45.3    & 74.8     \\
$\langle \tau/\tau_0 \rangle$ &
0.344    & 0.372    & 0.392    & 0.368    & 0.384    & 0.505    \\ 
\hline 
\hline
\end{tabular}
\caption{Summary  of  the  results  obtained  for  the  simulation  of
	microlenses towards the  LMC for the EROS model  for an age of
	the  halo  of  14~Gyr,   different  model  IMFs,  and  several
	magnitude cuts.}
\end{table*}

In  a second set  of Monte  Carlo calculations  we have  simulated the
observational data obtained by the EROS team.  We recall here that the
EROS collaboration  have not found any microlensing  event towards the
LMC and one candidate event towards the SMC.  Adopting a standard halo
model  and  assuming  $\tau_{\rm  SMC}=1.4\tau_{\rm  LMC}$,  the  EROS
results imply  an optical depth  $\tau_0=0.36\times10^{-7}$ (Tisserand
et al.  2006),  which is four times smaller than  that obtained by the
MACHO  team. Although it  is expected  that the  value of  the optical
depth obtained from our simulations should be only slightly different,
it is as  well true that this may  be a test of the  robustness of our
numerical procedures. In particular,  the detection efficiency of both
experiments is very different.  Additionally the areas (and the number
of objects)  surveyed by both  teams are different.  The  data results
are summarized in  table 3. Our simulations show  that the white dwarf
population could  account for a $35\%$  of the optical  depth found by
the EROS  team if a standard  initial mass function  is adopted, while
for the  non--standard initial mass  function the contribution  of the
white dwarf population could be as large as $50\%$.  On the other hand
the  expected number  of  objects has  an  upper limit  of  1 for  the
standard initial mass function and  2 for the log--normal initial mass
function. Both results  are in agreement with the  results of the EROS
experiment.  Again, as it was the case for the simulation of the MACHO
experiment, the  contribution of  ONe white dwarfs  is small.   All in
all,  it seems  that the  microlensing optical  depth obtained  by the
MACHO collaboration is a clear overestimate.

\begin{figure}
\vspace{13cm}
\includegraphics{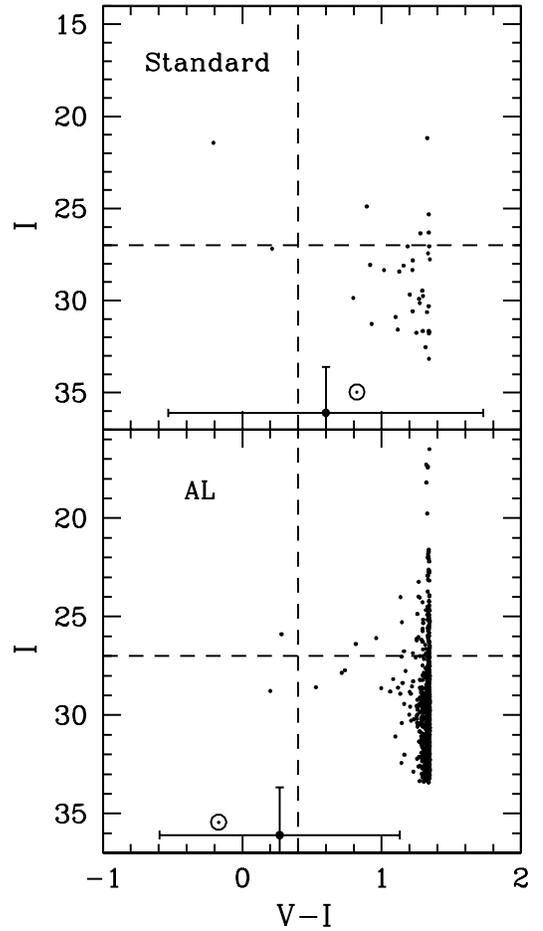} 
\caption{Color--magnitude  diagram for  the  white dwarf  distribution
         (ONe white dwarfs are circled)  for the HDF--S of two typical
         simulations.    The  dashed   line  respresents   the  HDF--S
         observation limit.  Also  represented is the average expected
         location within 1$\sigma$ error of a typical ONe white dwarf.
         See text for details.}
\end{figure}

\subsection{The Hubble Deep Field South}

Kilic et al.  (2005) have  recently re--observed the Hubble Deep field
south  (HDF--S), and  have found  three white  dwarf  candidates among
several  faint blue  objects which  exhibit significant  proper motion
and,  thus, are  assumed  to belong  to  the thick--disk  or the  halo
populations.  Consequently,  we have also performed a  series of Monte
Carlo simulations  in the direction of  the HDF--S $(l=328.25^{\circ}\
b=-49.21^{\circ})$ for  a small  window of $4.062\  {\rm arcmin}^{2}$.
We have used the Johnson--Cousins  $UBVRI$ system instead of the WFPC2
photometry because the differences between both photometric systems is
smaller  than $0.02^{\rm  mag}$ for  the range  of colors  under study
(Holtzman  et al.   1995).   Also,  no reddening  was  applied to  the
synthetic white dwarf stars. Contrary  to what has been done until now
the  results presented  in  this  section are  the  average of  $10^3$
different   realizations.   Each  of   these  realizations   has  been
normalized to  the local  density of halo  white dwarfs  as previously
described.  The synthetic white  dwarf population using this procedure
is shown in  the color--magnitude diagram of Fig.   4.  In this figure
we  represent  two  typical  simulations  for  the  halo  white  dwarf
population in  the direction  of the HDF--S  for the two  initial mass
functions under  study.  As  can be seen,  the number of  white dwarfs
susceptible to  be detected  in the HDF--S  survey --- that  is, those
with $I$  magnitude smaller than  $27^{\rm mag}$ ---  is substantially
larger for the log--normal initial  mass function of Adams \& Laughlin
(1996) than for the standard  initial mass function. Specifically, the
average  number of  objects  with  $I<27^{\rm mag}$  turns  out to  be
$6\pm2$  for the  case in  which a  standar initial  mass  function is
adopted, while  for the log--normal initial mass  function this number
is  $110\pm8$.  However,  and in  order to  avoid confusion  with blue
extragalactic objects  and main sequence  stars, Kilic et  al.  (2005)
restricted their search  for white dwarfs candidates to  colors in the
range  $V-I<0.4$.  Adding  this  new restriction  we  obtain that  the
expected number  of white  dwarfs should be  $1\pm1$ for  both initial
mass functions.   Although this result implies that  both initial mass
functions  are  compatible  with  the  observations,  the  log--normal
initial mass  function produces  a large number  of white  dwarfs with
colors  in  the interval  $0.6<V-I<1.4$,  which  has no  observational
counterpart.  Additionally, in  Fig.  4 we also show  the only one ONe
white dwarf obtained for each one of these two typical simulations. In
both  cases  its  location  is  shown  as  an  encircled  dot  in  the
color--magnitude diagram. It  is worth mentioning that in  most of the
$10^3$ realizations an ONe white dwarf is found, and thus we also show
the  average location  of  ONe white  dwarfs  in the  color--magnitude
diagram, along with the  corresponding $1\sigma$ error bars. Note that
in any case ONe white dwarfs are much fainter and bluer than normal CO
white dwarfs, as it should be expected given that for a typical age of
the halo most  ONe white dwarfs have already reached  the blue hook in
the color--magnitude diagram.

\subsection{The dark matter density}

The results discussed so far indicate  that, even in the case in which
the contribution  of ONe  white dwarfs is  taken into account,  only a
small fraction of  the microlensing optical depth towards  the LMC can
be  attributed  to white  dwarfs.   We recall  that  if  we adopt  the
microlensing optical  depth of the MACHO  experiment this contribution
is nearly  a $20\%$ for the  biased initial mass function  of Adams \&
Laughlin  (1996)  and  $\sim  10\%$  for  the  standard  initial  mass
function.   Besides,  for  a   spherical  isothermal  halo  model  the
microlensing optical depth towards the  LMC is given by the expression
(Alcock et al 2000; Griest 1991):

\begin{equation}
\tau_{\rm LMC}=5.1\times 10^{-7}f
\end{equation}

\noindent where $f$  is the fraction of the halo mass  that is made of
lensing objects.   Thus, the  white dwarf population  would contribute
$f\approx 0.05$ to the mass of the halo in the most optimistic case.

\begin{table}
\centering
\begin{tabular}{lcccc}
\hline      
\hline      
\multicolumn{1}{c}{\ }           &      
\multicolumn{2}{c}{Standard}     &   
\multicolumn{2}{c}{AL}           \\
\hline 
           &  CO  & ONe & CO & ONe    \\  
\cline{2-3}  
\cline{4-5}    
ISM &  $1.1\times10^{-4}$ &  $6.4\times10^{-6}$ & $2.8\times10^{-3}$ & $3.5\times10^{-5}$  \\   
WD  &  $5.4\times10^{-5}$ &  $9.5\times10^{-7}$ & $9.2\times10^{-4}$ & $5.3\times10^{-6}$  \\  
\hline
\hline
\end{tabular}
\caption{Density  of   baryonic  matter  $(M_{\sun}/$pc$^3)$   in  the
         Galactic  halo  within 300 pc from  the  Sun in  the form gas 
         returned to  the interstellar medium (ISM) and in the form of 
         white dwarfs (WD).}
\end{table}

However, we  can go  one step  beyond using the  results of  our Monte
Carlo  simulations. In particular,  we can  compute the  baryonic dark
matter density in  the form of white dwarfs  using the $1/V_{\rm max}$
method.   We  proceed as  follows.  For each  star  of  the sample  we
determine the maximum volume over  which each star can contribute as a
microlensing event using the expression

\begin{equation}
V_{\rm max}=\frac{\Omega}{3}(r_{\rm max}^3-r_{\rm min}^3)$$
\end{equation}

\noindent where $r_{\rm max}$ is the  radius of the volume in which we
distribute the objects of our sample,  which in our case is the radius
of Galactic halo, and $r_{\rm min}$  is the minimum volume for which a
white  dwarf still  belongs  to the  sample  considering its  apparent
magnitude to  be fainter  than the adopted  magnitude cut.   Then, the
number density of white dwarfs is

\begin{equation}
n=\sum_{i=1}^{N_{\rm obj}}\frac{1}{V_{{\rm max}_i}}.
\end{equation}

Using this procedure we find  that the contribution of white dwarfs to
the baryonic dark matter would be roughly a $3\%$ in the case in which
a standard intial  mass function is considered and  nearly a $5\%$ for
the  case in  which the  initial mass  function of  Adams  \& Laughlin
(1996) is adopted.

Finally, from our Monte Carlo  simulations we can also derive the {\sl
total} density of  baryonic matter in the Galactic  halo within 300 pc
from the Sun in the form  of main sequence stars, stellar remmants and
in  the  corresponding  ejected  mass.   We  obtain  $\rho_0=2.6\times
10^{-4}\, M_{\sun}$~pc$^{-3}$  for the standard  initial mass function
and  $3.8\times  10^{-3}\,  M_{\sun}$~pc$^{-3}$  for  the  log--normal
intial  mass function.   The respective  contributions of  CO  and ONe
white  dwarfs to  the mass  budget  and of  the mass  returned to  the
interstellar medium  are also shown in  Table 4.  Note  that the total
contribution of ONe white dwarfs is rather limited.  The total density
of baryonic  matter obtained from  our Monte Carlo simulations  can be
compared as well with the local dynamical matter density:

\begin{equation}
\rho_{\rm DM}=\frac{v_{\rm rot}^2}{4\pi{\rm G} R_{\sun}^2},
\end{equation}

\noindent where $v_{\rm  rot}$ is the rotation velocity  of the Galaxy
and  $R_{\sun}$ is  the  Galactocentric distance.  Thus, the  fraction
$\eta$ of baryonic matter of the Galaxy resulting from the white dwarf
population can  be estimated.  Our results indicate  that $\eta$ would
be  a modest  0.02  for the  case  in which  a  standard initial  mass
function  is adopted,  whereas  a sizeable  fraction  of the  baryonic
matter, $\eta=0.52$, can be accounted  if the initial mass function of
Adams \& Laughlin (1996) is assumed.

\section{Conclusions}

We have  analyzed the  contribution of ONe  white dwarfs to  the MACHO
content of the  Galactic halo. We find that  although ONe white dwarfs
fade to  invisibility very rapidly  and, thus, they are  good baryonic
dark matter candidates, their contribution to the microlensing optical
depth towards the LMC is  rather limited. In particular, we have found
that when the  contribution of ONe white dwarfs  is taken into account
the  microlensing  optical  depth  does  not  increase  significantly,
independently   of  the   adopted  initial   mass  function.   If  the
microlensing  optical  depth  is  adopted  to be  that  of  the  MACHO
experiment, $\tau_0=1.2\times 10^{-7}$ (Alcock  et al. 2000) --- which
probably  is an  overestimate ---  we find  that the  fraction  of the
microlensing optical depth due to  the whole white dwarf population is
at  most $\sim  13\%$ in  the case  in which  a standard  initial mass
function is  adopted and $\sim  19\%$ if the log--normal  initial mass
function of Adams \& Laughlin  (1996) is considered.  These values are
roughly  $\sim  3\%$  larger  than  those already  found  by  Garc\'\i
a--Berro et  al. (2004),  who only considered  the contribution  of CO
white dwarfs.   We have  also studied if  some of the  candidate white
dwarfs of the Hubble Deep Field South could be ONe white dwarfs and we
have found that  most probably this is not the  case. Finally, we have
also discussed the contribution of the whole white dwarf population to
the mass of the Galactic halo. We have found that this contribution is
of the order of a modest 5\%  in the most optimistic case. All in all,
we conclude that white dwarfs  are not significant contributors to the
mass of the Galactic halo.


\begin{acknowledgements}
Part   of    this   work   was    supported   by   the    MEC   grants
AYA05--08013--C03--01 and 02, by the European Union FEDER funds and by
the AGAUR.
\end{acknowledgements}



\begin{thebibliography}{1}

\bibitem{AL96}  Adams, F.C., \& Laughlin, G., 1996, \apj, 468, 686
\bibitem{Aea95} Alcock, C., Allsman,  R.A., Alves, D.,  Axelrod, T.S., 
                Bennett, D.P., Cook,  K.H., Freeman, K.C., Griest, K.,
                Guern,  J., Lehner,  M.J.,  Marshall, S.L.,  Peterson,
                B.A., Pratt, M.R., Quinn, P.J., Rodgers, A.W., Stubbs,
                C.W., \& Sutherland, W., 1995, ApJ, 454, L125
\bibitem{AEA97} Alcock,  C., Allsman, R.A., Alves,  D., Axelrod, T.S.,
		Becker,  A.C.,  Bennett,  D.P., Cook,  K.H.,  Freeman,
		K.C.,  Griest, K., Guern,  J., Lehner,  M.J., Mashall,
		S.L.,  Peterson,  B.A.,   Pratt,  M.R.,  Quinn,  P.J.,
		Rodgers, A.W., Stubbs, C.W., Sutherland, W., \& Welch,
		D.L., 1997, \apj, 486, 69
\bibitem{AEA00} Alcock, C., Allsman, R.A., Alves, D.R., Axelrod, T.S.,
		Becker,  A.C., Bennett, D.P.,  Cook, K.H.,  Dalal, N.,
		Drake,  A.J.,  Freeman, K.C.,  Geha,  M., Griest,  K.,
		Lehner,  M.J., Marshall,  S.L.,  Minniti, D.,  Nelson,
		C.A.,  Peterson,  B.A.,  Popowski,  P.,  Pratt,  M.R.,
		Quinn,  P.J., Stubbs,  C.W., Sutherland,  W., Tomaney,
		A.B., Vandehei, T., \& Welch, D., 2000, \apj, 542, 281
\bibitem{A06}   Althaus,  L.G.,  Garc\'\i   a--Berro,  E.,   Isern,  J.,
                C\'orsico,  A.H.,   \&  Rohrmann,  R.D.,   2007, \aap,
                465, 249
\bibitem{BEA05} Becker,  A.C.  Rest, A.  Stubbs,  C., Miknaitis, G.A.,
                Miceli, A.,  Covarrubias, R., Hawley,  S.L., Aguilera,
                C.,  Smith, R.C., Suntzeff,  N.B., Olsen,  K., Prieto,
                J.L., Hiriart, R., Garg,  A., Welch, D.L., Cook, K.H.,
                Nikolaev,  S., Clocchiatti,  A., Minniti,  D., Keller,
                S.C.\& Schmidt, B.P., 2005, IAU Symposium, 225, 357
\bibitem{BEL03} Belokurov,  V., Evans, N.W., Le Du,  Y., 2003, \mnras,
                341, 1373
\bibitem{BEL04} Belokurov,  V., Evans, N.W., Le Du,  Y., 2004, \mnras,
                352, 233
\bibitem{B87}   Binney,  J.,  \&  Tremaine,  H., 1987,  {\sl  Galactic
		Dynamics} (Princeton: Princeton Univ.  Press)
\bibitem{C97}   Canal,  R.,  Isern,  J.,  \& Ruiz--Lapuente, P., 1997, 
                \apj, 488, L35
\bibitem{Ch96}  Chabrier,  G.,  Segretain,  L.,  \&  M\'era, D., 1996, 
                \apj, 468, 21
\bibitem{Ch04}  Chabrier, G., 2004, \apj, 611, 315
\bibitem{CH95}  Charlot, S., \& Silk, J., 1995, \apj, 445, 124
\bibitem{Cea03} Cutri,  R.M., Skrutskie, M.F., van  Dyk, S., Beichman,
	        C.A.,  Carpenter,  J.M.,  Chester, T.,  Cambresy,  L.,
	        Evans, T., Fowler, J.,  Gizis, J., Howard, E., Huchra,
	        J.,  Jarrett,  T.,  Kopan,  E.L.,  Kirkpatrick,  J.D.,
	        Light, R.M., Marsh, K.A., McCallon, H., Schneider, S.,
	        Stiening, R., Sykes,  M., Weinberg, M., Wheaton, W.A.,
	        Wheelock, S.,  \& Zacarias,  N., 2003, {\sl  2MASS All
	        Sky Catalog of point sources}, Univ.  of Massachusetts
	        and IPAC/California Institute of Technology
\bibitem{DB98}  Dehnen, W.  \& Binney, J., 1998, \mnras, 298, 387
\bibitem{E06}   Eisenstein, D.J., Liebert, J., Harris, H.C., Kleinman,
	        S.J.,  Nitta,  A.,   Silvestri,  N.,  Anderson,  S.A.,
	        Barentine,  J.C.,  Brewington,  H.J.,  Brinkmann,  J.,
	        Harvanek, M., Krzesi{\'n}ski, J., Neilsen, E.H., Long,
	        D.,  Schneider, D.P., \&  Snedden, S.A.,  2006, \apjs,
	        167, 40
\bibitem{Fea97} Finley,  D.S.,  Koester,  D., \&  Basri, G., 1997, ApJ, 
                488, 375
\bibitem{GEA04} Garc\'\i  a--Berro, E.,  Torres,  S.,  Isern,  J.,  \&
		Burkert, A., 2004, \aap, 418, 53
\bibitem{GG01}  Gates, E.I., \& Gyuk, G., 2001, \apj, 547, 786
\bibitem{GEA02} Goldman,  B., Afonso,  A., Alard, Ch.,  Albert, J.-N.,
		Amadon, A.,  Andersen, J., Ansari,  R., Aubourg, \'E.,
		Bareyre, P.,  Bauer, F., Beaulieu,  J.-Ph., Blanc, G.,
		Bouquet,  A.,  Char,  S.,  Charlot, X.,  Couchot,  F.,
		Coutures, Ch., Derue,  F., Ferlet, R., Fouqu{\'e}, P.,
		Glicenstein,  J.-F., Gould, A.,  Graff, D.,  Gros, M.,
		Ha\"{\i}ssinski, J.,  Hamadache, C., Hamilton, J.-Ch.,
		Hardin,  D.,  Kat,  J.~de,  Kim,  A.,  Lasserre,  Th.,
		Le~Guillou, L., Lesquoy, {\'E}., Loup, C., Magneville,
		Ch., Mansoux,  B., Marquette, J.-B.,  Maurice, {\'E}.,
		Maury,     A.,    Milsztajn,    A.,     Moniez,    M.,
		Palanque-Delabrouille, N.,  Perdereau, O., Pr{\'e}vot,
		L., Regnault, N., Rich,  J., Spiro, M., Tisserand, P.,
		Vidal-Madjar,  A.,  Vigroux,  L., \&  Zylberajch,  S.,
		2002, \aap, 389, 69
\bibitem{G91}   Griest, K., 1991, \apj, 366, 412
\bibitem{GDG00} Gyuk, G.,  Dalal, N., \& Griest, K.,  2000, \apj, 535,
                90
\bibitem{H01}   Hambly,  N.C.,  Irwin,   M.J., \&  MacGillivray, H.T., 
                2001, \mnras, 326, 1295
\bibitem{H98}   Hansen, B.M.S., 1998, \nat, 394, 860
\bibitem{HEA06} Holopainen,  J., Flynn, C., Knebe, A.,  Gill, S.P., \&
                Gibson, B.K., 2006, \mnras 368, 1209
\bibitem{HEA95} Holtzman, J.A.,  Burrows, C.J., Casertano, S., Hester,
                J.J., Trauger, J.T.,  Watson, A.M., Worthey, G., 1995,
                \pasp, 107, 1065
\bibitem{IEA99} Ibata, R.A., Richer,  H.B., Gilliland, R.L., \& Scott,
		D., 1999, \apj, 524, L95
\bibitem{I89a}  Iben, I., \& Laughlin, G., 1989, \apj, 341, 312
\bibitem{I98}   Isern,  J.,   Garc\'\i  a--Berro,   E.,   Hernanz,  M.
                Mochkovitch, R., \& Torres, S., 1998, \apj, 503, 239
\bibitem{J90}   James, F., 1990, Comput.  Phys.  Commun., 60, 329
\bibitem{K06}   Kallivayalil, N., van  der  Marel,  R.P., Alcock,  C.,
                Axelrod, T., Cook, K.H.,  Drake, A.J., Geha, M., 2006,
                \apj, 638, 772
\bibitem{KE98}  Kerins, E., Evans N.W., 1998, \apj 503, 75
\bibitem{K05}   Kilic, M., Mendez, R.A.,  Von Hippel,  T.,  \& Winget,
                D.E., 2005, \apj, 633, 1126
\bibitem{K02}   Klypin, A., Zhao, H., \&  Somerville, R.,  2002, \apj,
                573, 597
\bibitem{LEA01} Lasserre, T.  Afonso.  C., Albert, J.N., Andersen, J.,
		Ansari, R.,  Aubourg, {\' E}., Bareyre, P., Bauer, F.,
		Beaulieu,   J.P., Blanc,  G., Bouquet,  A.,  Char, S.,
		Charlot,   X., Couchot, F.,  Coutures, C.,  Derue, F.,
		Ferlet,   R., Glicenstein,  J.F., Goldman,  B., Gould,
		A.,  Graff,  D., Gros,  M., Haissinski,  J., Hamilton,
		J.C., Hardin,   D., de Kat, J., Kim,  A., Lesquoy, {\'
		E}.,   Loup,   C.,   Magneville,   C.,  Mansoux,   B.,
		Marquette,   J.B., Maurice,  {\'  E}., Milsztajn,  A.,
		Moniez, M.,  Palanque-Delabrouille, N., Perdereau, O.,
		Pr{\' e}vot,   L., Regnault, N., Rich,  J., Spiro, M.,
		Vidal-Madjar,   A., Vigroux,  L.,  \& Zylberajch,  S.,
		2001, \aap, 355, L39
\bibitem{LBH05} Liebert, J., Bergeron, P., \& Holberg, J., 2005, ApJS, 
                156, 47
\bibitem{LDM89} Liebert, J., Dahn, C.C., \& Monet, D.G., 1989, in {\sl
                ``White Dwarfs''},  Ed. G. Wegner  (Berlin: Springer),
                15
\bibitem{MEA99} Muraki, Y.,  Sumi, T., Abe, F., Bond,  I., Carter, B.,
                Dodd,  R.,  Fujimoto, M.,  Hearnshaw,  J., Honda,  M.,
                Jugaku,  J.,  Kabe,   S.,  Kato,  Y.,  Kobayashi,  M.,
                Koribalski, B., Kilmartin,  P., Masuda, K., Matsubara,
                Y., Nakamura, T., Noda, S., Pennycook, G., Rattenbury,
                N.,  Reid,   M.,  Saito,  T.,  Sato,   H.,  Sato,  S.,
                Sekiguchi, M.,  Sullivan, D., Takeuti,  M., Watase Y.,
                Yanagisawa,  T.,  Yock,  P.  \& Yoshizawa,  M.,  1999,
                Progress of Theoretical Physics Supplement, 133, 233
\bibitem{NFW}   Navarro, J.F.,  Frenck, C.S., \& White,  S.D.M., 1997,
                \apj, 490, 493
\bibitem{P86}   Paczy\'nski, B., 1986, \apj, 304, 1
\bibitem{R05}   Rahvar, S., 2005, \mnras, 356, 1127
\bibitem{S94}   Sahu, K., 1994, Nature, 370, 275
\bibitem{SEA00} Salaris,   M.,  Garc\'{\i}a--Berro, E.,  Hernanz,  M.,
		Isern, J.  \& Saumon, D., 2000, \apj, 544, 1036
\bibitem{S98}   Scalo, J., 1998,  in  {\sl  The  Stellar Initial  Mass
		Function},  Eds.:  G.   Gilmore  \&  D.   Howell  (San
		Francisco: PASP Conference Series), Vol.  142, 201
\bibitem{S68}   Schmidt, M., 1968, \apj, 151, 393
\bibitem{S07}   Silk, J., 2007, in {\sl ``The invisible universe: dark
                matter and dark energy''},  Proc.  of the Third Aegean
                Summer  School (Berlin:  Springer  Verlag), in  press,
                {\tt astro-ph/0603209}
\bibitem{TEA06} Tisserand,  P.,  Le Guillou,  L., Afonso,  C., Albert,
                J. N., Andersen, J., Ansari, R., Aubourg, E., Bareyre,
                P.,  Beaulieu,  J.  P.,  Charlot,  X.,  Coutures,  C.,
                Ferlet, R., Fouqu\'e, P., Glicenstein, J. F., Goldman,
                B., Gould,  A., Graff,  D., Gros, M.,  Haissinski, J.,
                Hamadache, C., de Kat,  J., Lasserre, T., Lesquoy, E.,
                Loup, C.,  Magneville, C., Marquette,  J. B., Maurice,
                E.,   Maury,   A.,    Milsztajn,   A.,   Moniez,   M.,
                Palanque-Delabrouille,   N.,  Perdereau,   O.,  Rahal,
                Y. R., Rich, J., Spiro, M., Vidal-Madjar, A., Vigroux,
                L., \& Zylberajch, S., 2006, {\tt astro-ph/0607207}
\bibitem{TEA98} Torres,  S., Garc\'\i  a--Berro,  E.,  \&  Isern,  J.,
		1998, \apj, 508, L71
\bibitem{UEA94} Udalski,  A., Szymanski, M., Kaluzny,  J., Kubiak, M.,
                Mateo, M., \& Krzeminski,  W., 1994, Acta Astron., 44,
                1
\bibitem{Vea02} Vennes,  S.,  Smith,  R.J.,  Boyle,  J., Croom,  S.M.,
                Kawka,  A.,  Shanks, T.,  Miller, L., \&  Loaring, N.,
                2002, \mnras, 335, 673
\bibitem{Z98}   Zhao, H.S., 1998, \mnras, 294, 139

\end{thebibliography}
\end{document}